\documentclass[aps,prd,showpacs,twocolumn,superscriptaddress,amsmath,amsfont,amssymb,graphicx]{revtex4}
\usepackage{color,graphicx,epsfig} \usepackage{ifpdf}
\usepackage{amsmath} \usepackage{amssymb} \usepackage{bm}
\usepackage{color}

\usepackage[english]{babel}

\newcommand {\E}[1]{\times 10^{#1}}	
\newcommand {\e}[1]{\mathrm{~#1}}       

\definecolor{Red}{rgb}{1.,0.,0.}

\begin{document}

\title{Can scalar leptoquarks explain the $f_{D_s}$ puzzle?}

\author{Ilja Dor\v sner} 
\email[Electronic address:]{ilja.dorsner@ijs.si}
\affiliation{J. Stefan Institute, Jamova 39, P. O. Box 3000, 1001 Ljubljana, Slovenia}
\affiliation{Faculty of Natural Sciences, Zmaja od Bosne 33-35, 
71000 Sarajevo, Bosnia and Herzegovina}

\author{Svjetlana Fajfer} 
\email[Electronic address:]{svjetlana.fajfer@ijs.si} 
\affiliation{J. Stefan Institute, Jamova 39, P. O. Box 3000, 1001 Ljubljana, Slovenia}
\affiliation{Department of Physics,
  University of Ljubljana, Jadranska 19, 1000 Ljubljana, Slovenia}

\author{Jernej F. Kamenik} 
\email[Electronic address:]{jernej.kamenik@ijs.si} 
\affiliation{J. Stefan Institute, Jamova 39, P. O. Box 3000, 1001
  Ljubljana, Slovenia}
\affiliation{INFN, Laboratori Nazionali di Frascati, Via E. Fermi 40 I-00044 Frascati, Italy}

\author{Nejc Ko\v{s}nik} 
\email[Electronic address:]{nejc.kosnik@ijs.si} 
\affiliation{J. Stefan Institute, Jamova 39, P. O. Box 3000, 1001
  Ljubljana, Slovenia}

\date{\today}

\begin{abstract}
  Motivated by the disagreement between experimental and lattice QCD
  results on the $D_s$ decay constant we systematically reinvestigate
  role of leptoquarks in charm meson decays. We consider scalar
  leptoquarks that transform as a weak interaction triplet, doublet, or
  singlet in a model independent approach, and also argue that in a
  particular $SU(5)$ GUT model these leptoquark states, contained in
  the $45$-dimensional Higgs representation, could be safe against
  proton decay bounds. Using the current experimental measurements in
  $\tau$, kaon and charm sectors, we find that scalar leptoquarks
  cannot naturally explain the $D_s \to \mu \nu$ and $D_s \to \tau
  \nu$ decay widths simultaneously. While any contributions of the
  triplet leptoquarks are already excluded, the singlets could only
  contribute significantly to the $D_s\to\tau\nu$ width. Finally, a
  moderate improvement of the experimental upper bound on the
  $D^0\to\mu^+\mu^-$ decay width could exclude the doublet
  contribution to the $D_s\to\mu\nu$, while present experimental data
  limits its mass to be below $1.4\e{TeV}$. Possible new signatures at present and near future
  experiments are also briefly discussed.
\end{abstract}


\maketitle

\section{Introduction}
Leptoquark states are expected to exist in various extensions of the
Standard Model~(SM). They were first introduced in the early grand
unification theories (GUTs) in the
seventies~\cite{Georgi:1974sy,Pati:1974yy}. Scalar leptoquarks are
expected to exist at TeV scale in extended technicolor models as well
as in models of quark and lepton compositeness. Scalar quarks in
supersymmetric models with R-parity violation (RPV) may also have
leptoquark-type Yukawa couplings~\footnote{For a recent review
  c.f. pages 452-455 in \cite{Amsler:2008zzb}}.

Recently, discrepancies between the experimental measurements of
leptonic decay modes of $D_s$
mesons~\cite{:2007ws,:2008sq,Alexander:2009ux,Onyisi:2009th} and the
lattice results for the relevant $f_{D_s}$ decay
constant~\cite{Follana:2007uv, Bernard:2009wr, Blossier:2009bx} have
stimulated many analyses.  One intriguing indication is that the
central measured and predicted values for the $f_{D_s}$ differ by more
than $10\,\%$ with a combined significance of $D_s\to\tau\nu$ and
$D_s\to\mu\nu$ channels of roughly $2.3\,\sigma$~\cite{JabseyCharm09},
while the corresponding values for $f_D$ are in perfect agreement. In
Ref.~\cite{Dobrescu:2008er} the idea of scalar leptoquarks has been
revived to explain the missing decay widths. Some implications of this
suggestion have been further explored using
semileptonic~\cite{Benbrik:2008ik, Kronfeld:2008gu} and rare charm
decays~\cite{Fajfer:2008tm}.

Generally, leptoquarks which also couple to diquarks mediate fast
proton decay and are therefore required to be much above the
electroweak scale~\cite{Weinberg:1980bf}, making them uninteresting
for other low energy phenomena.  ``Genuine'' leptoquarks on the other
hand, couple only to pairs of quarks and leptons, and may thus be
inert with respect to proton decay. In such cases, proton decay bounds
would not apply and leptoquarks may produce signatures in other
low-energy phenomena.  In this article we set out to study whether
scalar leptoquarks can naturally account for the $f_{D_s}$ puzzle and
at the same time comply with all other measured flavor observables.

We consider all possible renormalizable leptoquark interactions with
SM matter fields consistent with the SM gauge symmetry.  One can
construct such dimension-four operators using leptoquarks which are
either singlets, doublets or triplets under the $SU(2)_L$. If we
furthermore require that such leptoquarks contribute to leptonic
decays of charged mesons at tree level, we are left with three
possible representation assignments for the $SU(3)_c\times
SU(2)_L\times U(1)_Y$ gauge groups: $(\mathbf 3, \mathbf 3,-1/3)$,
$({\mathbf {\bar 3}}, \mathbf{2},-7/6)$ and
$(\bm{3},\bm{1},-1/3)$. Only the weak doublet leptoquark is
``genuine'' in the above sense. However, using a concrete $SU(5)$ GUT
model where the relevant leptoquarks are embedded into the
45-dimensional Higgs representation~($\bm{45}_H$), we demonstrate how
leptoquark couplings to matter can arise and in particular, how the
dangerous couplings to diquarks -- both direct and
indirect~\cite{Weinberg:1980bf} -- can be avoided.

\section{General considerations}

In our analysis we will assume the mass eigenstates within a
leptoquark weak multiplet to be nearly degenerate.  While
large mass splittings within a weak multiplet may be
considered unnatural, more importantly, they are also tightly
constrained by electroweak precision observable
$T$~\cite{Keith:1997fv}. Consequently, one generically gets
correlations between semileptonic charged currents and (lepton) flavor
violating neutral currents, which represent important constraints on
any leptoquark scenario trying to resolve the $D_s$ leptonic widths
puzzle. Also, we focus on observables mediated by the relevant
leptoquark couplings at tree level since these already involve
processes forbidden in the SM at tree level, i.e., flavor changing
neutral currents (FCNCs) and lepton flavor violation (LFV)
processes. Finally, since the present $f_{D_s}$ deviation is of mild
significance, we require all the measured constraints to be satisfied
within one standard deviation (at $68\,\%$ C.L.) except upper bounds, for
which we use published $90\,\%$ C.L. limits. We consider a leptoquark
explanation of the $f_{D_s}$ discrepancy as natural, if both $D_s$ and
$D$ leptonic decay widths can be obtained close to their measured
central values.

After the electroweak~(EW) symmetry breaking, quarks and leptons
acquire their masses from their respective Yukawa interactions. These
induce in the physical (mass eigen-)basis a CKM and PMNS rotations
between the upper and the lower components of the fermion
doublets. Consequently, it is impossible to completely isolate
leptoquark mediated charged current interactions to a particular quark
or lepton generation in the left-handed sector {\it irrespective of
  the initial form of the leptoquark couplings to SM matter fields,
  unless there is some special alignment with the right-handed quark
  sector}.  To see this, we denote as $X^{(}{}'{}^{)}$ a $3\times 3$
arbitrary Yukawa matrix in the weak basis, and write down flavor
structure of interaction of the quark and lepton doublet parts
\begin{subequations}
\label{eq:FlavStruct}
\begin{align}
  \overline{Q^w_q} X^{q \ell} =& ( \overline{u^w_q} \quad \overline{
    d^w_q})  X^{q \ell}
  = (\overline{u_q} \quad \overline{d'_q}) (U^\dagger X)^{q \ell},\\
  X'^{q \ell } L^w_\ell =& X'^{q \ell} (\nu^w_\ell \quad
    e^w_\ell)^T = (X' E)^{q \ell} (
    \nu'_\ell \quad e_\ell )^T,
\end{align}
\end{subequations}
where fields with $w$ superscript are in the weak basis, whereas $d' =
V_{CKM} d$ and $\nu' = V_{PMNS} \nu$. The unitary matrices $U, D, E$,
and $N$ rotate the fields from mass to weak basis and are unphysical
\emph{per se}, so we absorb them in redefinition of the couplings
(e.g. $Y_{LQ} \equiv U^\dagger X$ on the quark and $Y_{LQ}' \equiv X' E$ on the
lepton side) and consider them as free parameters. In what follows, we will use the
convention where all the remaining rotations are assigned
to down-type quark ($V_{CKM} = U^\dagger D$) and neutrino ($V_{PMNS} =
E^\dagger N$) sectors, and the quark mass-eigenstates are defined as
\begin{equation*}
(Q_1, Q_2, Q_3) = \begin{pmatrix}
u & c & t\\
d' & s' & b'
\end{pmatrix}, \ \
 \begin{pmatrix}d'&s'&b'\end{pmatrix}
= \begin{pmatrix}d&s&b\end{pmatrix} V^T_\mathrm{CKM}.
\end{equation*}
It is obvious in this notation that, even if the $Y_{LQ}$ matrix had all rows, 
except for the $q$-th one, set to zero, which would correspond to leptoquark
coupling \emph{only} to $u_q$, one would still get non-zero couplings to all three left-handed down-quarks. Same rationale
holds true for the lepton sector due to $V_{PMNS}$, but since the neutrino
flavors are not tagged in present experiments, the respective decay
widths are summed over all neutrino flavors. Whenever a
mass-eigenstate antineutrino $\bar \nu_i$ is produced in a reaction,
its amplitude includes, according to Eq.~(\ref{eq:FlavStruct}), a
factor of $\sum_j Y_{LQ}'^{qj} V_{PMNS}^{ji}$ for leptoquark
interaction, or $V_{PMNS}^{l i}$ if the neutrino was produced in $W\ell
\nu$ vertex. In any case, when one sums the rates for all neutrino
species
\begin{equation*}
  \sum_{i=1,2,3} |\mathcal A_i|^2 \sim \sum_{i=1,2,3} V_{PMNS}^{ji} V_{PMNS}^{li*} = \delta^{jl},
\end{equation*}
it becomes evident that in the summed rate all the neutrino indices
are replaced by the lepton flavors. This is equivalent to the absence
of mixing in the lepton doublets. 

The above considerations are more general and similar in spirit to the
ones recently discussed in Ref.~\cite{Blum:2009sk} for the case of
$K-\bar K$ and $D -\bar D$ mixing. In fact, any new physics coupling
to SM fermionic weak doublets exhibits similar kind of correlations,
and contributions to charged current transitions cannot be isolated to
a particular quark generation.

Another important particularity of the $f_{D_s}$ puzzle is that it
affects a Cabibbo favored $c\to s$ transition. Consequently, the
hierarchy of correlations with other processes is largely determined
by the CKM mixing hierarchy. In particular, the mixing of the third
generation with the first two is much smaller than the mixing of the
first two generations among each other. Therefore, for our purposes,
it is often a good approximation, to completely neglect effects of the
third generation in the quark sector. Then we can parameterize a
generic leptoquark coupling in the weak basis using a common
(real) prefactor and a rotation angle 
\begin{equation*}
Y_{LQ}^{q\ell} = y^{\ell}_{LQ} (\sin\phi, \cos\phi).
\end{equation*}
In addition, the only CKM rotation is due to the Cabibbo angle and
there is no SM CP violating phase ($d' = \cos \theta_c d + \sin
\theta_c s, s' = -\sin \theta_c d + \cos \theta_c s$ and
$V_{us}=-V_{cd}=\sin\theta_c=0.225$).  The absence of SM phases is not
critical for our purposes, since we only consider CP conserving
quantities, and since the relevant SM amplitudes in our considered
processes have approximately the same weak phase even in the full
three generation case. The leptoquark couplings themselves, however,
could in principle have arbitrary new phases. These could be important
in processes with two or more interfering amplitudes contributing, at
least one of those being due to the leptoquarks. We deal with this
possibility on a case by case basis. Finally, in all the scenarios
considered we have checked explicitly that the two generation
approximation is valid by performing numerical leptoquark parameter
scans including the full CKM structure and a full set of possible
leptoquark couplings with arbitrary phases. In this case,
(semi)leptonic $B$ decays $B\to\tau\nu$ and especially $B\to D\tau\nu$
can be used to put additional constraints on the leptoquark parameters
relevant to the $D_s\to\tau\nu$ width. Numerically however, these
constraints turn out not be competitive with the others due to
presently limited experimental precision.

\section{Triplet leptoquark $(\mathbf 3, \mathbf 3,-1/3)$}

The triplet leptoquark can in principle couple to diquarks and thus destabilize the proton, so
one has to check in an underlying model if that is indeed the
case. The allowed leptoquark interaction
Lagrangian consists of a single term
\begin{equation}
  \mathcal L_3 = Y_3^{ij}\, \overline{Q_i^{c}} i \tau_2\, \bm\tau\cdot \bm\Delta_3^* \, L_j + \mathrm{h.c.} \,,
\label{eq:L3}
\end{equation}
where $\overline{Q^c} = -Q^T C^{-1}$, $C = i \gamma^2 \gamma^0$ and
$\bm\tau$ are the Pauli matrices. The $3\times 3$ coupling matrix
$Y_3$ is arbitrary in the bottom-up approach.  On the other hand, its
entries may be related to other parameters in an UV embedding of the
effective theory. In the concrete $SU(5)$ model analyzed in the
Appendix~\ref{app}, the above couplings are due to the contraction of
$\bm{10}$ and $\bm{\overline{5}}$ with $\bm{45}^*_H$ -- also
responsible for giving masses to the down quarks and charged
leptons. A different contraction of $\bm{10}$ and $\bm{10}$ with
$\bm{45}_H$ couples the triplet to diquarks.
The latter term can be consistently set to zero in the supersymmetric
version of the model, thus sufficiently suppressing proton decay.

As already mentioned in Ref.~\cite{Dobrescu:2008er}, the triplet
leptoquarks cannot by themselves account for deviations in
$D_s\to\ell\nu$ for both $\tau$ and $\mu$ in the final state due to
constraints coming from LFV tau decays, such as $\tau \to
\eta^{(}{}'{}^{)} \mu$ and $\tau\to\phi\mu$. Numerically, the $\tau
\to \eta \mu$ decay turns out to be most constraining. The triplet
leptoquark contribution can be written as
\begin{equation*}
  \Gamma^{(3)}_{\tau\to\eta\mu} = \frac{\left|\sum_{q=u,d,s} \chi_q \tilde Y^{q\tau}_3 \tilde Y_3^{q\mu*} {f^q_\eta}\right|^2  }{512 \pi  m_{\Delta_3}^4} m_\tau^3 \left[1-\left(\frac{m_\eta}{m_\tau}\right)^2\right]^{3/2}\,,
\end{equation*}
where we have neglected the muon mass. The weight $\chi_q = 1$ is for
$q=u$ and $2$ for $q=d,s$ comes from an additional $\sqrt{2}$ factor in
the interaction terms with $\Delta_3 (t_3 = \pm 1)$ states. Decay
constants of $\eta$ meson $f_\eta^q$ are defined as
in~\cite{Feldmann:1998vh}. As explained in the previous section, the couplings $\tilde Y_3$ contain an
additional $V_{CKM}$ rotation for the down-type quarks
\begin{equation*}
  \tilde Y_3^{q\ell} \equiv \left\{\begin{array}{ccc}
      Y_3^{q\ell} &;& q = u,c,t,\\
      (V_{CKM}^T Y_3)^{q\ell} &;& q = d,s,b.
    \end{array} \right.
\end{equation*}
Thus, the upper bound on $\tau \to \eta \mu$ decay width directly
constrains the products $\tilde Y_3^{s\tau} \tilde Y_3^{s\mu*}$,  $\tilde Y_3^{d\tau} \tilde Y_3^{d\mu*}$ and $Y_3^{u\tau}
Y_3^{u\mu*}$(=$(\cos\theta_c\tilde
Y^{d\tau}_3+\sin\theta_c \tilde Y^{s\tau}_3)(\cos\theta_c\tilde
Y^{d\mu*}_3+\sin\theta_c \tilde Y^{s\mu*}_3)$ in the two-generations approximation). On the other hand, the relative contributions to the $D_s$
leptonic widths~(see Eq.~(\ref{eq:3Dsrelative})) in the two
generations approximation are $\tilde Y_3^{s\mu}  (\tilde
Y^{s\mu*}_3-\tan\theta_c \tilde Y^{d\mu*}_3)$ for
the muon channel and $\tilde Y_3^{s\tau} (\tilde
Y^{s\tau*}_3-\tan\theta_c \tilde Y^{d\tau*}_3)$ for the tau
channel. Both cannot be sizable and at the same time agree with the
bounds that come from the $\tau \to \eta \mu$ decay width \footnote{Phases of $Y_3^{ij}$ could in principle conspire to yield a small contribution to $\Gamma_{\tau\to\eta\mu}$, however, such solutions cannot at the same time reproduce other related LFV decay limits, such as $\tau\to\eta'\mu$, $\tau\to\pi\mu$ or $\tau\to\phi\mu$.}. In scenario of
triplet leptoquarks therefore, {\it one of the measured leptonic
  channels $D_s \to \ell \nu$ would necessarily have to be a
  measurement artifact}. We will consider both possibilities
separately.

If the leptoquarks have sizable coupling $\tilde Y_3^{s\tau}$
(implying $\tilde Y_3^{s\mu} \sim 0$ by the $\tau \to \eta \mu$ decay width)
we can obtain a non-zero contribution to the $D_s \to \tau \nu$ decay
width due to the interference term between the SM and the leptoquark
amplitude in
\begin{equation*}
  \Gamma^{(3)}_{D_s\to\tau\nu} = \Gamma^{SM}_{D_s\to\tau\nu} \left|1 + \frac{\delta_3^\tau}{4\sqrt{2} G_F } \right|^2\,,
\end{equation*}
where the SM width is
\begin{equation*}
\Gamma^{SM}_{D_s\to\tau\nu} = \frac{G_F^2 m_\tau^2\left|V_{\text{cs}}\right|{}^2 f_{D_s}^2 m_{D_s}}{8 \pi }  \left[1-\left(\frac{m_\tau}{m_{D_s}}\right)^2\right]^2,
\end{equation*}
and the relative contribution of the triplet leptoquarks reads
\begin{equation}
\delta_3^\tau \equiv \frac{Y_3^{c\tau*} \tilde Y_3^{s\tau}}{V_{cs} m_{\Delta_3}^2}.
\label{eq:3Dsrelative}
\end{equation}
An important observation is that the leptoquarks in this scenario
contribute to the same effective operator as the SM and thus exhibit
the same helicity suppression. In the two generations approximation,
the relative triplet contribution simplifies to $\delta_3^{\tau} =
(y_3^{\tau})^2 \cos\phi \sin\phi (\tan\theta_c +
\cot\phi)/m_{\Delta_3}^2$. Reproducing the measured branching ratio
$Br(D_s\to\tau\nu)=0.0561(44)$~\cite{JabseyCharm09} while using the
most precise lattice input
$f_{D_s}=241(3)\e{MeV}$~\cite{Follana:2007uv} would require
$\sqrt{\delta_3^{\tau}} \approx 0.002\e{GeV^{-1}}$. On the other hand
the $Y_3^{u\tau}$ coupling of leptoquarks
is constrained by precise measurement of the lepton flavor
universality ratio $\pi_{\mu/\tau} \equiv Br(\tau\to\pi\nu)/Br(\pi\to\mu\nu) =
0.1092(7)$~\cite{Amsler:2008zzb,Pich:2008ni}.
Leptoquarks contribute to semileptonic tau decays in the form 
\begin{equation}
\Gamma^{(3)}_{\tau\to\pi\nu} = \Gamma^{SM}_{\tau\to\pi\nu} \times  \left|1 + \frac{1}{4\sqrt{2} G_F } 
  \left[\frac{Y_3^{u\tau*} \tilde Y_3^{d\tau}}{V_{ud} m_{\Delta_3}^2}\right]\right|^2,
  \label{eq:tauSemileptonic}
\end{equation}
where
\begin{equation*}
  \Gamma^{SM}_{\tau\to\pi\nu} = \frac{G_F^2 m_\tau^3 \left|V_{\text{ud}}\right|{}^2 f_{\pi}^2 }{16 \pi }  \left[1-\left(\frac{m_\pi}{m_\tau}\right)^2\right]^2\,.
\end{equation*}
In the two generations approximation the term in square brackets can
be written as $(y_3^{\tau})^2 \sin\phi \cos\phi (-\tan\theta_c +
\tan\phi)/m_{\Delta_3}^2$. To exactly satisfy both the
$\pi_{\mu/\tau}$ value and explain leptonic $D_s \to \tau \nu$ excess
we need either
\begin{itemize}
\item[(a)] $\tan\phi \approx \tan\theta_c$, i.e., leptoquarks couple
  only to $s$ but not to $d$ quark ($\tilde Y_3^{d\tau} \approx 0$), or
\item[(b)]
$\sin\phi \approx 0$, i.e., leptoquarks couple only to
$c$ but not to $u$ quark ($Y_3^{u\tau}=0$).
\end{itemize}

\begin{figure}[t]
\begin{center}
\begin{tabular}{c}
\includegraphics[width=7cm]{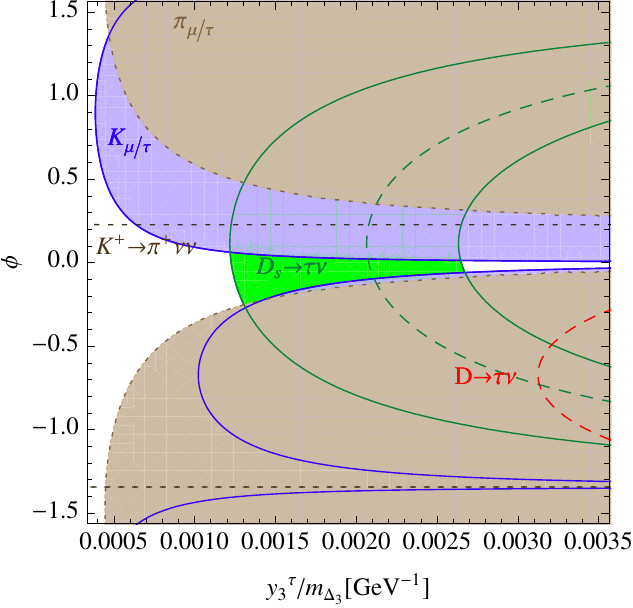}\\
\\
\includegraphics[width=7cm]{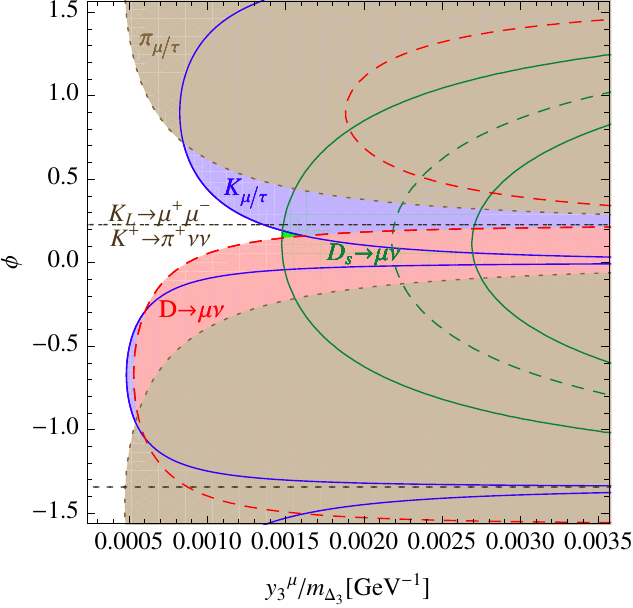}
\end{tabular}
\end{center}
\caption{ Combined bounds on the triplet leptoquark parameters in the
  two-generation limit in the tau (upper plot) and muon (lower plot)
  sectors. All bands represent $68\%$ C.L. exclusion intervals, except
  the upper bound on $D\to\tau\nu$ which is taken at $90\%$ C.L.. The
  $K^+\to\pi^+\nu\bar\nu$ and $K_L\to\mu^+\mu^-$ constraints can only
  be satisfied on the two horizontal dashed lines. Within the green
  bands, the $D_s\to\ell\nu$ excess can be accounted
  for. \label{fig:1} }
\end{figure}

However, in the limit (a) one must have a sizable coupling
$Y_3^{u\tau}$ due to CKM rotation which results in relative
contribution of size $\delta_3^{\tau}$ to the Cabibbo suppressed
semileptonic tau decays $\tau \to K \nu$ (of the form
(\ref{eq:tauSemileptonic}) with appropriate flavor replacement $d\to
s$ and $\pi\to K$).  These are measured in agreement with the SM at
the $3\,\%$ level~\footnote{At this level of precision the main
  theoretical uncertainty in the SM comes from electromagnetic
  corrections, which have to be taken into
  account~\cite{Cirigliano:2007ga}.} (in particular the ratio $K_{\mu/\tau} \equiv Br(\tau
\to K \nu)/Br(K \to \mu \nu) = 0.0109(4)$
\cite{Amsler:2008zzb,Pich:2008ni})~\footnote{A complementary triplet
  leptoquark contribution to the rare decay $D^0\to\mu^+\mu^-$
  presently gives a weaker constraint and is not included in Fig.~\ref{fig:1}}.

In the other limit, (b), one must have sizable coupling $\tilde
Y_3^{d\tau}$ and thus gets a relative contribution scaling as
$\delta_3^{\tau} $ to the $D\to \tau \nu$ decay width. Currently only
an upper bound exists for this channel $Br(D\to\tau\nu)<1.2\E{-3}$ at
$90\,\%$ C.L.~\cite{:2008sq}. Even more importantly, one gets
a non-vanishing contribution to the rare $K^+\to\pi^+\nu\bar\nu$
decay. Since the triplet leptoquark contributes with the same
effective operator as the SM, its contribution can be obtained by
simply replacing the $\lambda_t X_t$ product in the master formula
of~\cite{Mescia:2007kn} with
\begin{equation*}
  \lambda_t X_t \to \lambda_t X_t + \frac{\sqrt{2}\pi \sum \tilde Y_3^{s\ell} \tilde Y_3^{d\ell*}}{{G_F} \alpha_{em}  \sin\theta_W m_{\Delta_3}^2}\,.
\end{equation*}
This process is measured to have
$Br(K^+\to\pi^+\nu\bar\nu)=(17.3^{+11.5}_{-10.5})\E{-11}$~\cite{Adler:2001xv,
  Artamonov:2008qb}. It constraints the sum of leptoquark coupling
combinations $\tilde Y_3^{s\ell} \tilde Y_3^{d\ell*} = (y_3^{\ell})^2
\cos\theta_c\cos\phi(\tan\phi - \tan\theta_c)(1+\tan\theta_c\tan\phi)$
and fixes very accurately $\tan\phi=\tan\theta_c$ or
$\tan\phi=-\cot\theta_c$. The constraint applies to all lepton flavors
since it is inclusive with respect to neutrino flavor  \footnote{Possible phases in $Y_3^{ij}$ cannot invalidate this bound, since tau LFV decays require the $Y_3^{i\ell}$ couplings to different $\ell$ lepton flavors to be of different orders of magnitude.}. The combined
impact of all these constraints on the triplet leptoquark contribution
to $D_s\to\tau\nu$ is shown on the first plot in Fig.~\ref{fig:1}.
One observes that the combination of the strong bounds coming from
$K^+\to\pi^+\nu\bar\nu$ combined with $Br(\tau \to K \nu)/Br(K \to \mu
\nu)$ completely excludes the triplet leptoquarks from explaining the
$D_s\to\tau\nu$ excess.

In the opposite scenario where the leptoquarks couple to muons
the situation is similar to the tau case with two differences: (1) the
$D\to\mu\nu$ decay width has already been measured and the
$Br(D\to\mu\nu)=3.8(4)\E{-4}$~\cite{:2008sq} agrees perfectly with the SM
prediction using the most precise lattice QCD value of
$f_D=208(4)\e{MeV}$~\cite{Follana:2007uv}; (2) an additional
constraint comes from the FCNC decay $K_L\to\mu^+\mu^-$ as it receives
contributions from leptoquarks of the form
\begin{equation*}
  \Gamma^{(3)}_{K_L\to\mu^+\mu^-} = \frac{f_K^2 m_K m_{\mu}^2 \sqrt{1-\frac{4m_{\mu}^2}{m_K^2}}}{64 \pi} \frac{\left|\tilde Y_3^{s\mu} \tilde Y_3^{d\mu*}\right|^2 }{m_{\Delta_3}^4}\,.
\end{equation*}
The requirement that such leptoquark contributions do not exceed
presently measured $Br(K_L\to\mu^+\mu^-) = 6.84(11)
\E{-9}$~\cite{Amsler:2008zzb} produces a bound equivalent to the
existing one coming from $K^+\to\pi^+\nu\bar\nu$. Combining these two
additional constraints with the rest also clearly disfavors a triplet
leptoquark explanation of the $D_s\to\mu\nu$ excess, as shown on
the bottom plot in Fig.~\ref{fig:1}.

\section{Doublet leptoquark $({\mathbf {\bar 3}}, \mathbf{2},-7/6)$}

The doublet leptoquarks are innocuous as far as proton decay is
concerned. The allowed dimension four interactions in this case are
\begin{equation}
  \mathcal L_2 = Y_{2L}^{ij}\,  \overline Q_i \,i \tau_2 \Delta_2^*\, e_j  
+ Y_{2R}^{ij}\, {\overline u_i}  \Delta_2^\dagger L_j  + \mathrm{h.c.}\,.
\label{eq:L2}
\end{equation}
In the particular $SU(5)$ model, the term proportional to $Y_{2R}$ stems
from the contraction of $\bm{10}$ and $\bm{\overline{5}}$ with
$\bm{45}^*_H$ while the $Y_{2L}$ term is due to $\bm{10}$
and $\bm{10}$ being contracted with $\bm{45}_H$.

\begin{figure}[t]
\begin{center}
\includegraphics[width=7cm]{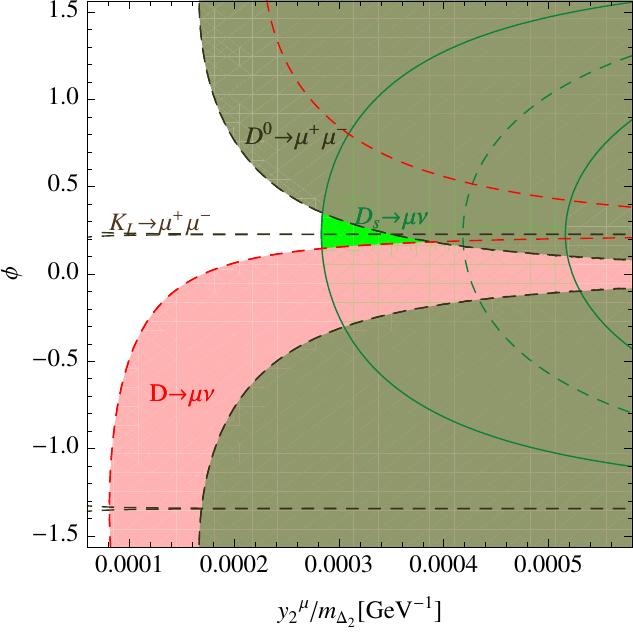}
\end{center}
\caption{ Combined $D^0\to\mu^+\mu^-$ and $D\to\mu\nu$ bounds on the doublet leptoquark parameters in the two-generation limit in muon sector as explained in the text. $D\to\mu\nu$ band represent $68\%$ C.L. exclusion interval, while the upper bound on $D^0\to\mu^+\mu^-$  is taken at $90\%$ C.L.. The $K_L\to\mu^+\mu^-$ constraint can only
  be satisfied on the two horizontal dashed lines. Within the green band, the $D_s\to\ell\nu$ excess can be accounted for. \label{fig:2} }
\end{figure}

In this scenario the same states couple left-handed quarks to
right-handed leptons and vice versa.  Consequently, only the product
of both couplings can contribute to the $D_s$ leptonic widths through
the interference with the SM
\begin{align}
  \label{eq:2semilep}
  \Gamma^{(2)}_{D_s\to\ell\nu} &= \Gamma^{SM}_{D_s\to\ell\nu} \left| 1
    -
    \frac{\delta_2^\ell}{4\sqrt{2} G_F}\right|^2,\\
  \delta_2^\ell &\equiv \frac{m_{D_s}^2}{m_\ell (m_c + m_s)}
  \frac{Y_{2R}^{c\ell*} \tilde Y_{2L}^{s\ell}} {V_{cs}^*
    m_{\Delta_2}^2}\,\nonumber.
\end{align}
Again, the couplings of left-handed down- and up-type quarks are
misaligned
\begin{equation*}
  \tilde Y_{2L}^{q\ell} \equiv (V_{CKM}^\dagger Y_{2L})^{q\ell} \textrm{ for } q = d,s,b.
\end{equation*}
Note that the doublet leptoquark contribution exhibits no helicity
suppression. Thus, explaining both muon and tau leptonic partial
widths of $D_s$ requires vastly different leptoquark
couplings.  On the other hand, now one also has to take into account
the strict bound coming from the decay $D^0 \to \mu^+\mu^-$.  In the
doublet leptoquark model, this mode receives potential contributions
from several coupling combinations
\begin{align}
  \Gamma^{}_{D^0 \to \mu^- \mu^+} = &\frac{f_D^2 m_D^3}{512 \pi}
  \,\sqrt{1- \frac{4m_\mu^2}{m_D^2}} \left[
    \left(1-\frac{4m_\mu^2}{m_D^2} \right) A + B\right],
  \label{eq:D0mumu}
\end{align}
where $A$ and $B^{}$ contain the couplings of doublet leptoquarks
\begin{align*}
  A = &\frac{m_D^2}{m_{\Delta_2}^4 (m_c+m_u)^2} \left|
    Y_{2R}^{u\mu} Y_{2L}^{c\mu*} -
      Y_{2R}^{c\mu*} Y_{2L}^{u\mu}\right|^2,\\
  B = &\frac{1}{m_{\Delta_2}^4} \left| \frac{m_D}{m_c+m_u}
    \left(Y_{2R}^{u\mu} Y_{2L}^{c\mu*} +
      Y_{2R}^{c\mu*} Y_{2L}^{u\mu}\right)\right.\\
  &\left.\phantom{\frac{1}{m_c}}+ \frac{2 m_\mu}{m_D} \left(Y_{2R}^{c\mu*} Y_{2R}^{u\mu} +
      Y_{2L}^{c\mu*} Y_{2L}^{u\mu}\right)\right|^2.
\end{align*}
Two combinations involve $Y_{2R}^{u\mu}$ which is in conjunction with
$\tilde Y_{2L}^{s\mu}$ and $\tilde Y_{2L}^{s\tau}$ constrained through
precision kaon and tau lepton flavor universality tests similarly as
in the triplet scenario.  In addition, this coupling does not
contribute to the $D_s\to \mu\nu$ width~(\ref{eq:2semilep}). The
remaining two combinations can be rewritten by using the Cabibbo
rotation in terms of $Y_{2R}^{c\mu} Y_{2L}^{u\mu*} = Y_{2R}^{c\mu} (
\cos\theta_c \tilde Y_{2L}^{d\mu*} + \sin\theta_c \tilde
Y_{2L}^{s\mu*} )$ and $Y_{2L}^{c\mu} Y_{2L}^{u\mu*} = ( \cos\theta_c
\tilde Y_{2L}^{s\mu} - \sin\theta_c \tilde Y_{2L}^{d\mu} ) (
\cos\theta_c \tilde Y_{2L}^{d\mu*} + \sin\theta_c \tilde
Y_{2L}^{s\mu*} ) $.  The $D\to\mu\nu$ width measurement constrains
directly the size of $\tilde Y_{2L}^{d\mu}$. In absence of this coupling,
$D^0\to\mu^+\mu^-$ would receive dominant contributions from just two
non-interfering leptoquark amplitudes
\begin{align*}
  A \approx &\frac{m_D^2}{m_{\Delta_2}^4 (m_c+m_u)^2} \left|
    Y_{2R}^{c\mu} \tilde Y_{2L}^{s\mu*} \sin\theta_c \right|^2,\\
  B \approx &\frac{\sin^2 \theta_c}{m_{\Delta_2}^4} \left| \frac{m_D}{m_c+m_u}
    Y_{2R}^{c\mu*} \tilde Y_{2L}^{s\mu}
    + \frac{2 m_\mu}{m_D} \left(\tilde Y_{2L}^{s\mu}\right)^2 \cos\theta_c \right|^2\,,
\end{align*}
where $A$ can be related to $D_s \to \mu\nu$ decay width contribution
(it is proportional to $\sin^2\theta_c \cos^2\theta_c
|\delta_2^{\mu}|^2$).  The term proportional to $A$ alone yields
$Br(D^0\to\mu^+\mu^-)\approx 8.3\E{-7}$ for the central value of
$D_s\to\mu\nu$ decay width. Recently, an improved experimental limit
of $Br(D^0\to\mu^+\mu^-) < 4.3\E{-7}$ at $90\,\%$ C.L. was put forward
by CDF~\cite{CDF9226}.  It is evident that this introduces some
tension between explaining the $D_s$ excess and not spoiling the
agreement in the $D$ case. Due to the moderate significance of the
$D_s$ discrepancy, this tension is not yet conclusive as can be seen
on Fig.~\ref{fig:2}, where we plot the combined constraints in the
$\phi$--$y_2^\mu$ plane. Fig.~\ref{fig:2} is generated in the
following way. We first parameterize $\tilde Y_{2L}^{s\mu} =
y_{2L}^\mu \cos\phi $ and $\tilde Y_{2L}^{d\mu} = y_{2L}^\mu \sin\phi
$. We then set $Y_{2R}^{u\mu}=0$ while $Y_{2R}^{c\mu} =
y_{2R}$. Finally, we vary $y^{\mu}_{2L}$ and $y^{\mu}_{2R}$ while
keeping the product $y^{\mu}_2 = \sqrt{y^{\mu}_{2L}y^{\mu}_{2R}}$
fixed and use the best fit value to determine the allowed region.  We
include the $Br(K_L\to\mu^+\mu^-)$ constraint which is also relevant
in this case
\begin{equation*}
  \Gamma^{(2)}_{K_L\to\mu^+\mu^-} = \frac{\left|\tilde Y_{2L}^{s\mu} \tilde Y_{2L}^{d\mu*}\right|^2 f_K^2 m_K^5 \sqrt{1-\frac{4 m_{\mu }^2}{m_K^2}} \left(1-\frac{2 m_{\mu }^2}{m_K^2}\right)}{256 \pi  m_{\Delta _2}^4 \left(m_d+ m_s\right)^2}\,.
\end{equation*}
While both central values for $D$ and $D_s$ leptonic widths clearly
cannot be reproduced by the doublet leptoquark contribution, a future
improvement of the bound on $D^0\to\mu^+\mu^-$ is clearly sought after
to reach a definite conclusion on this scenario.

Opposed to the triplet leptoquark case, the verdict on the
$D_s\to\tau\nu$ contribution of the doublet leptoquark is still far
from conclusive. Firstly, because the corresponding $D$ leptonic mode
has not been measured. Secondly, because there are presently no
strong experimental bounds on FCNCs in the up quark sector involving
only tau leptons or only neutrinos (doublet leptoquark does not
contribute to $s\to d \nu\bar\nu$ transitions). We note in passing
that, provided the doublet leptoquarks are to explain both tau and
muon final state excesses, there is an important bound coming from the
aforementioned $\tau\to \eta^{(}{}'{}^{)}\mu$ decays. In this scenario
they constrain the following combination of parameters
$\delta_2^{LFV}=|Y_{2L}^{s\tau}Y_{2L}^{*s\mu}|/m_{\Delta_2}^2$
appearing in
\begin{equation*}
  \Gamma^{(2 s)}_{\tau\to\eta\mu} = \frac{\left|\tilde Y_{2L}^{s\tau} \tilde Y_{2L}^{s\mu*}\right|^2 m_\tau (f^s_{\eta })^2 m_\eta^4 \left(1-\frac{m_{\eta }^2}{m_\tau^2}\right)^2}{128 \pi  m_s^2 m_{\Delta_2 }^4}\,,
\end{equation*}
where we have only considered the $s$-quark contribution and have
neglected the muon mass. On the other hand, explanation of $D_s$
leptonic excesses requires nonzero values for $\delta_2^{\mu,\tau}$.
Finally, in order not to spoil perturbative treatment of the
couplings, none of the couplings should exceed a value of roughly
$|Y_{2L,R}^{ij}| < \sqrt{4\pi}$. Then, one can combine the above inequalities
to yield a robust {\it upper bound on the doublet leptoquark mass}:
\begin{equation*}
  m_{\Delta_2} < \sqrt{4\pi\,{\delta_2^{LFV}}/{|\delta_2^{\mu}\delta_2^{\tau}|}}\,.
\end{equation*}
Taking the present $Br(\tau\to\eta\mu)<6.5\E{-8}$ at $90\,\%$
C.L. bound~\cite{Amsler:2008zzb} and the central values for the two
leptonic decay widths, one obtains a value of roughly $1.4\e{TeV}$,
which is certainly within the LHC reach~\cite{Mitsou:2004hm}.

\section{Singlet leptoquark $(\bm{3},\bm{1},-1/3)$}

This state was originally proposed to explain the $D_s$ leptonic width
puzzle in Ref.~\cite{Dobrescu:2008er}. On the other hand, singlet
leptoquarks are notorious for their mediation of proton
decay. However, as in the case of the leptoquark triplet, one can
demonstrate in concrete $SU(5)$ embedding (see Appendix~\ref{app}) that the
dangerous couplings do not necessarily appear. Similarly to triplets,
weak singlets can also couple to pairs of SM matter weak
doublets. However, now also couplings to pairs of singlets are
possible resulting in the dimension four interaction Lagrangian with
two terms
\begin{equation}
  \mathcal L_1 = Y_{1L}^{ij} \overline{Q_i^c} i \tau_2 \Delta_1^* L_j  + Y_{1R}^{ij} \overline{u_i^c}  \Delta_1^* e_j + \mathrm{h.c.}\,.
  \label{eq:L1}
\end{equation}
In this generic effective theory description clear correlations among
different charged and neutral current flavor observables, present in
the triplet case, are somewhat diluted by the presence of the second
interaction term which modifies the singlet leptoquark contribution to
the $D_s$ leptonic width
\begin{align}
  &\frac{\Gamma_{D_s\to\tau\nu}^{(1)}}{\Gamma_{D_s\to\tau\nu}^{SM}} =  \Bigg | 1
  +
  \frac{1}{4\sqrt{2} G_F m_{\Delta_1}^2 } \nonumber\\
  &\times \left\{ \left[\frac{Y_{1L}^{c\tau*} \tilde
        Y_{1L}^{s\tau}}{V_{cs} }\right] - \frac{ m_{D_s}^2
      \left(Y_{1R}^{c\tau*} \tilde Y_{1L}^{s\tau} \right)}{ V_{cs}^*
      m_\tau (m_c + m_s) } \right\}\Bigg|^2\,,
  \label{eq:tripletDsLeptonic}
\end{align}
where $\tilde Y^{q\ell}$ are defined as in the triplet leptoquark
scenario. The second term in Eq.~(\ref{eq:L1}) can come from the
$SU(5)$ embedding without causing any conflict with the bounds on
proton decay lifetime even if the leptoquark is very light, whereas
the presence of the first term would require some fine tuning in order
for the leptoquark not to couple to diquarks (see Appendix~\ref{app}).
Note that if the first term is absent, then the singlet leptoquark
cannot contribute to the $D_s$ leptonic decay width. If the
second term is absent the analysis is analogous to the triplet
leptoquark scenario, with the exception that the singlet does not
contribute to $K_L\to\mu^+\mu^-$.  Such is for example the case of the RPV  minimal supersymmetric SM, where the interaction term of a right-handed down squark to quark and lepton doublets is present and corresponds to the first term in (\ref{eq:L1}), while the second term is absent. 

From the triplet scenario we know
that $K^+\to\pi^+\nu\bar\nu$ forces the $Y^\ell_{1L}$ couplings to be
diagonal in the down-type quark basis and in particular $\tilde
Y_{1L}^{d\ell}\approx 0$.  Also relevant is the constraint from the
lepton flavor universality ratio $Br(\tau\to K\nu)/Br(K\to\mu\nu)$
which receives relative leptoquark contributions of the form
(\ref{eq:tripletDsLeptonic}) with suitable flavor replacement ($c\to
u$), while the tau semileptonic width is given as
\begin{align*}
  \frac{\Gamma^{(1)}_{\tau\to K\nu}}{ \Gamma^{SM}_{\tau\to K\nu}} = \Bigg|&1 + \frac{1}{4\sqrt{2} G_F }\\
  &\times\left\{ \left[\frac{Y_{1L}^{u\tau*} \tilde
        Y_{1L}^{s\tau}}{V_{us} }\right] - \frac{ m_\tau
      \left(Y_{1R}^{u\tau*} \tilde Y_{1L}^{s\tau} \right)}{ V_{us}^*
       (m_u + m_s) } \right\} 
  \Bigg|^2.
\end{align*}
Remaining constraint is the rare decay $D^0\to\mu^+\mu^-$ which in
this case is of the form (\ref{eq:D0mumu}) with $A$ and $B$
\begin{align*}
  A = &\frac{m_D^2}{m_{\Delta_1}^4 (m_c+m_u)^2} \left|
    Y_{1R}^{u\mu} Y_{1L}^{c\mu*} -
      Y_{1R}^{c\mu*} Y_{1L}^{u\mu}\right|^2,\\
  B = &\frac{1}{m_{\Delta_1}^4} \left| \frac{m_D}{ m_c+ m_u}
    \left(Y_{1R}^{u\mu} Y_{1L}^{c\mu*} +
      Y_{1R}^{c\mu*} Y_{1L}^{u\mu}\right)\right.\\
  &\left.\phantom{\frac{1}{\hat m_c}}- \frac{2 m_\mu}{m_D} \left(Y_{1R}^{c\mu*} Y_{1R}^{u\mu} +
      Y_{1L}^{c\mu*} Y_{1L}^{u\mu}\right)\right|^2.
\end{align*}
The remaining relevant free parameters can correspondingly be chosen as
an overall coupling magnitude $\delta$ and two angles ($\phi$,
$\omega$), defined through $\tilde Y_{1L}^{s\mu} = y_1^{\mu}
\sin\omega$, $Y^{c\mu}_{1R}=y_1^\mu \cos\omega\cos\phi$ and
$Y^{u\mu}_{1R}=y_1^\mu \cos\omega\sin\phi$. The value of $y_1^\mu$ is
bounded from above by the condition of perturbativity
($y_1^{\mu}<\sqrt{4\pi}$). Together with existing direct experimental
searches for second generation
leptoquarks~\cite{Abulencia:2005ua,Abazov:2006vc} this gives an
additional constraint on the possible size of the leptoquark
contributions to the $D_s$ leptonic width. By performing a numerical
fit of ($y_1^{\mu},\omega,\phi$) to these constraints we obtain the
result, that the experimental value for $Br(D_s\to\mu\nu)$ cannot be
reproduced within one standard deviation without violating any of the
other constraints, thus excluding the singlet leptoquark as a natural
explanation of the $D_s\to\mu\nu$ puzzle. 

As in the doublet case, the lack of experimental information on
up-quark FCNCs involving only tau leptons leaves the verdict on the
singlet leptoquark contribution to the $D_s\to\tau\nu$ decay width
open. What is certain is that due to the $K^+\to\pi^+\nu\bar\nu$
constraint any such contribution has to be aligned with the down-type
quark Yukawas such that $\tilde Y_1^{d\tau}\approx 0$ can be
ensured. In addition, the second term in Eq.~(\ref{eq:L1}) needs to be
present and sizable to avoid the bounds coming from
$Br(K\to\mu\nu)/Br(\tau\to K\nu)$.

\section{Conclusions}

Scalar leptoquarks cannot naturally explain both enhanced
$D_s\to\ell\nu$ decay widths due to existing constraints coming from
precision kaon, tau, and $D$ meson observables. The triplet leptoquark
is excluded from contributing to any of the widths. Sizable
contributions due to single right-handed down squark exchange in RPV
supersymmetric models are also excluded, while a generic leptoquark
singlet is definitely excluded only from explaining the $D_s\to\mu\nu$
width. The doublet contribution to this process is still technically
allowed, while an improvement in the search for $D^0\to\mu^+\mu^-$
could very soon also completely exclude it. For the $D_s\to\tau\nu$
only the triplet (and RPV) explanations are already excluded, while
the possible doublet explanation of both widths requires its mass to
lie below $1.4\e{TeV}$ and will certainly also be probed with direct
leptoquark production at the LHC. Possible future signatures of a
scenario where leptoquarks are responsible for the $D_s\to\tau\nu$
width could also be $Br(J/\psi\to\tau^+\tau^-)$ at the level of
$10^{-11}$, probably beyond the reach of BESIII~\cite{Asner:2008nq},
and also $Br(t\to c \tau^+ \tau^-)$ at the level of $10^{-5}$, close
to the limiting sensitivity of the LHC~\cite{Jana:2008kb}.

\begin{acknowledgments}
  This work is supported in part by the European Commission RTN
  network, Contract No. MRTN-CT-2006-035482 (FLAVIAnet), the Marie
  Curie International Incoming Fellowship within the $6^{th}$ European
  Community Framework Program (I.D.) and by the Slovenian Research
  Agency.
\end{acknowledgments}

\bibliography{refs}

\appendix

\section{$SU(5)$ EMBEDDING}
\label{app}
We now demonstrate i) how natural it is for the weak triplet, doublet 
and singlet leptoquark interaction terms to arise in renormalizable  
$SU(5)$ model, and ii) how plausible it is for them to be light enough to play 
role in flavor physics phenomena. 
 
 In $SU(5)$, an $i$th ($i=1,2,3$) generation of the SM matter fields
 comprises $\bm{10}_i(=(\bm{1},\bm{1},1)_i\oplus(\overline{\bm{3}},\bm{1},-2/3)_i
 \oplus(\bm{3},\bm{2},1/6)_i=(e^C_i,u^C_i,Q_i))$ and $\overline{\bm{5}}_i(=(\bm{1},\bm{2},-1/2)_i\oplus
 (\overline{\bm{3}},\bm{1},1/3)_i=(L_i,d^C_i))$, where $Q_i=(u_i \quad
 d_i)^T$ and $L_i=(\nu_i \quad e_i)^T$. The up quark (down quark and charged lepton) masses originate
 from the contraction of $\bm{10}_i$ and $\bm{10}_j$ ($\overline{\bm{5}}_j$) 
 with $5$- and/or $45$-dimensional Higgs representation. (Observe that
 $\bm{10} \times \bm{10} =
 \overline{\bm{5}}\oplus\overline{\bm{45}}\oplus\overline{\bm{50}}$  and $\bm{10}
 \times \overline{\bm{5}} = \bm{5}\oplus\bm{45}$.) Only 
 these two representations contain 
 component that is both electrically neutral
 and an $SU(3)_c$ singlet that can thus obtain phenomenologically allowed
 vacuum expectation value (VEV). Actually, both are needed in a realistic
 renormalizable setting on purely phenomenological grounds. 
  
 The most general renormalizable set of Yukawa coupling contractions with $\bm{5}_H$ and $\bm{45}_H$ is
 \begin{eqnarray*}
 \label{potential} V &=& Y_{5^*}^{ij}  \bm{10}^{\alpha \beta}_i
 \overline{\bm{5}}_{\alpha j} \bm{5}_{H \beta}^* +Y_5^{ij} \epsilon_{\alpha \beta \gamma \delta \epsilon}
   \bm{10}^{\alpha \beta}_i \bm{10}^{\gamma
 \delta}_j \bm{5}_H^\epsilon \nonumber\\
 &+& Y_{45^*}^{ij} 
 \bm{10}^{\alpha \beta}_i \overline{\bm{5}}_{\delta j}
 \bm{45}^{*\,\delta}_{H \alpha \beta}
  +
 Y_{45}^{ij} \epsilon_{\alpha \beta \gamma \delta \epsilon}
   \bm{10}^{\alpha
 \beta}_i \bm{10}^{\zeta \gamma}_j \bm{45}^{\delta
 \epsilon}_{H \zeta},
 \end{eqnarray*}
 where Greek indices are contracted in the $SU(5)$ space. Relevant
 fermion mass matrices are
 \begin{subequations}
 \label{M}
 \begin{align}
 \label{M_D}
 M_D =& \left( Y^T_{5^*} v^*_5  +  2  Y^T_{45^*}  v^*_{45}
 \right)/\sqrt{2} ,\\
 M_E =& \left( Y_{5^*}  v^*_5  - 6  Y_{45^*}  v^*_{45}
 \right)/\sqrt{2},
 \label{GJ}\\
 \label{M_U} M_U =& \left[ 4  (Y^T_{5}+Y_{5})  v_5  -   8 
 (Y^T_{45}-Y_{45})  v_{45}\right]/\sqrt{2},
 \end{align}
 \end{subequations}
 where $\langle\bm{5}_{H}^5\rangle=v_5/\sqrt{2}$,
 $\langle \bm{45}^{1 5}_{H 1} \rangle= \langle \bm{45}^{2
 5}_{H 2} \rangle=\langle \bm{45}^{3 5}_{H 3} \rangle =v_{45}/\sqrt{2}$ and
 $|v_5|^2+|v_{45}|^2=v^2$ ($v=247$\,GeV). $Y_{5^*}$, $Y_{45^*}$, $Y_{5}$ and
 $Y_{45}$ are arbitrary $3 \times 3$ Yukawa matrices. 
  
 If only $5$-dimensional ($45$-dimensional) Higgs representation
 were present one would have $M_E^T=(-3) M_D$. 
 A scenario with only one
 Higgs representation would hence yield $m_\tau/m_b=m_\mu/m_s=m_e/m_d$ at the
 GUT scale, which is in conflict with what is inferred from
 experimental observations. This is why both $\bm{5}_H$ and $\bm{45}_H$
 are needed at renormalizable level. (Note, since $m_\mu/m_s >1 $ 
 whereas $m_\tau/m_b \approx 1$ at the GUT scale this would suggest that the the $Y_{45^*}^{22}$ entry is
enhanced compared to other entries of $Y_{45^*}$~\cite{Georgi:1979df}.) 

Conveniently enough, the $45$-dimensional Higgs representation 
 $\bm{45}_H(=(\Delta_1,
 \Delta_2, \Delta_3, \Delta_4, \Delta_5, \Delta_6, \Delta_7) =
 (\bm{8},\bm{2},1/2)\oplus (\overline{\bm{6}},\bm{1}, -1/3) \oplus
 (\bm{3},\bm{3},-1/3) \oplus (\overline{\bm{3}}, \bm{2}, -7/6) \oplus
 (\bm{3},\bm{1}, -1/3) \oplus (\overline{\bm{3}}, \bm{1}, 4/3) \oplus
 (\bm{1}, \bm{2}, 1/2)$ contains a weak triplet ($\Delta_3$), 
 doublet ($\Delta_4$), and singlet ($\Delta_5$) leptoquarks we 
 are interested in whereas the $5$-dimensional Higgs representation, $\bm{5}_H$, 
 contains a singlet leptoquark only. So, these leptoquark states must be 
 present in any renormalizable theory based on $SU(5)$.

 The most
 stringent constraints on leptoquark masses and their couplings to matter 
 originate from limits on partial
 proton decay lifetimes. In that respect only $\Delta_4$ is
 innocuous enough since it does not directly mediate proton decay. (It cannot couple to a
 quark-quark pair.) It is also practically impossible for it
to be a part of the process that destabilizes proton through mixing with the 
Higgs doublet and some other state that couples to a quark-quark pair 
since $(\overline{\bm{3}},\bm{2}, 1/6)$ -- the only suitable candidate -- is not part of either $5$- or 
$45$-dimensional Higgs representation (or $24$-dimensional representation). 
It is thus phenomenologically possible for $\Delta_4$ to 
be light and have couplings to the matter fields of the form given in Eq.~(\ref{eq:L2}) in $SU(5)$. In that case 
$Y_{2L}=- 2^{1/2} [Y^\dagger_{45}-Y^*_{45}]$ and $Y_{2R}=Y^\dagger_{45^*} $.
 
It is also possible to have $\Delta_3$ that couples to the
quark-lepton pairs and no proton decay. In particular, the
$\bm{10}$-$\bm{\overline{5}}$-$\bm{45}^*_H$ contraction yields a
lepton-quark pair couplings with $\Delta_3$ of the form given in
Eq.~(\ref{eq:L3}): $Y_{3}=Y^\dagger_{45^*} $. On the other hand, the
$\bm{10}$-$\bm{10}$-$\bm{45}_H$ contraction yields couplings of
$\Delta_3$ to a quark-quark pair only.  Clearly, if only one of these
two possible contractions is present there would not be a tree level
proton decay due to $\Delta_3$.  In the former case there would not be
proton decay due to the mixing of $\Delta_3$ with the Higgs doublet
and some other states either since $(\bm{3},\bm{1}, 2/3)$ -- which
would be a suitable candidate -- is not part of either $5$- or
$45$-dimensional representation.

Finally, $\Delta_5$ could also be coupled to matter in a manner that renders proton stable contrary
to the usual expectation. Namely,
the $\bm{10}$-$\bm{10}$-$\bm{45}_H$ contraction yields couplings to a lepton-quark pair only. (This should be compared to the $\bm{10}$-$\bm{10}$-$\bm{5}_H$ contraction
that generates both the lepton-quark and quark-quark type of couplings simultaneously for the singlet leptoquark in $\bm{5}_H$.) This contraction yields the second term in 
Eq.~(\ref{eq:L1}): $Y_{1R}=2^{1/2}
[Y^\dagger_{45} -Y^*_{45}]$. The $\bm{10}$-$\bm{\overline{5}}$-$\bm{45}^*_H$ contraction, on the other hand, yields 
not only the second term in 
Eq.~(\ref{eq:L2}), i.e., $Y_{1L}=-2^{1/2}
Y^\dagger_{45^*}$, but also the quark-quark couplings which would lead to proton instability. If only $\bm{10}$-$\bm{10}$-$\bm{45}_H$ contraction is present proton could be stable and accordingly $\Delta_5$ could be light. 
Interestingly enough, 
it is possible to have a scenario in which there would not be any leptoquark induced proton decay. 
The necessary condition for this to happen would be the absence of the  
$\bm{10}$-$\bm{\overline{5}}$-$\bm{45}^*_H$ contraction.
\end{document}